\documentclass[11pt,a4paper]{article} 
\pdfoutput=1

\usepackage{jcapmod} 
\usepackage{booktabs}
\usepackage{url}

\newcommand{\eq}[1]{eq. (\ref{#1})} \newcommand{\fig}[1]{figure \ref{#1}}
\newcommand{\sektion}[1]{section \ref{#1}}
\newcommand{\subsektion}[1]{section \ref{#1}}
\newcommand{\ocite}[1]{ref. \cite{#1}}

 \newcommand{\ud}{\mathrm{d}}
\newcommand{\mchi}{m_{\chi}} \newcommand{\mphi}{m_{\phi}}
\newcommand{\gchi}{\Gamma_{\chi}} \newcommand{\gphi}{\Gamma_{\phi}}
\newcommand{\nphi}{N_{\phi}}
\newcommand{\pz}{\mathcal{P}_{\zeta}}

\title{Mixed inflaton and curvaton scenario with sneutrinos}

\author{Vedat Nefer \c{S}eno\u{g}uz}

\affiliation{Department of Physics, Do\u{g}u\c{s} University, 34722
Kad{\i}k\"oy, \.Istanbul, Turkey}

\emailAdd{nsenoguz@dogus.edu.tr}

\abstract{A variation of sneutrino inflation based on $\chi^2$ potential is
considered where the inflaton and the late-decaying field are sneutrinos of
different generations. The lighter, late-decaying sneutrino dilutes the
gravitinos over-produced after inflaton decay and generates the matter
asymmetry. It can also significantly contribute to the curvature perturbation,
realizing the mixed inflaton-curvaton case. The cosmic microwave background
(CMB) observables can distinguish this case from inflation with $\chi^2$
potential, provided that the initial value of the late-decaying sneutrino is either an
order of magnitude smaller or larger than the reduced Planck scale.}


\begin{document} \maketitle \flushbottom

\section{Introduction and review} \label{review}

Inflation \cite{Guth:1980zm} can be realized most simply by means of a scalar
field $\chi$, called the inflaton. In supersymmetric models, a well motivated inflaton
candidate is one of the sneutrinos, the scalar partners of the right-handed (RH)
neutrinos which determine the mass scale of the light neutrinos via the see-saw
mechanism \cite{Minkowski:1977sc}.  A minimal scenario where inflation is driven
by a sneutrino with $\chi^2$ potential works remarkably well
\cite{Murayama:1992ua,Ellis:2003sq}.

Among the attractive features of sneutrino inflation with $\chi^2$ potential
is that the sneutrino mass fixed from the amplitude of the primordial curvature
perturbation has the same order of magnitude with the see-saw scale inferred from
the light neutrino mass differences. Furthermore, this sneutrino mass ($\sim10^{13}$ GeV) is
also compatible with baryogenesis via leptogenesis \cite{Fukugita:1986hr},
with the lepton asymmetry originating from the decays of the inflaton-sneutrino
\cite{Hamaguchi:2001gw}. The predictions for
the spectral index $n_s$ and tensor to scalar ratio $r$ are compatible with
experiments including the WMAP results \cite{Komatsu:2010fb}.  

Sneutrino inflation predicts a reheat temperature $T_r$ of order $10^{14}$ GeV
if neutrino Yukawa couplings are of order unity (see \subsektion{sninf}). On the
other hand, to avoid the over-production of gravitinos in the early universe,
typically $T_r\lesssim10^9$ GeV is required
\cite{Khlopov:1984pf}.
Avoiding the gravitino problem thus requires very small Yukawa couplings, and as
a consequence the RH neutrino superpartner of the inflaton decouples from the
see-saw mechanism in the sneutrino inflation and leptogenesis scenario discussed
in refs. \cite{Murayama:1992ua,Ellis:2003sq}.

It is useful to consider variations of this scenario to interpret upcoming
experimental results, particularly those expected from the Planck mission
\cite{planck}. Here we do not require the RH neutrino superpartner of the
inflaton to have such small Yukawa couplings so that it decouples from the
see-saw mechanism. Instead, we assume that there is another, lighter generation
sneutrino with small Yukawa couplings, whose late decay can dilute the
gravitinos produced earlier. This lighter, late-decaying sneutrino can act as a
curvaton \cite{Lyth:2001nq,Lyth:2002my}, or can partially contribute to the
primordial curvature perturbation (the mixed inflaton-curvaton case
\cite{Linde:1996gt,Moroi:2001ct,Langlois:2004nn,Moroi:2005kz,Ichikawa:2008iq,Kobayashi:2012ba})
and also cause a second epoch of inflation
\cite{Starobinsky:1986fxa,Silk:1986vc} depending on its initial value. We show
that sufficient matter asymmetry can be generated while satisfying the gravitino
constraint in the mixed inflaton-curvaton case and work out the predictions
for the CMB observables.\footnote{For earlier work related to the curvaton
mechanism see refs.  \cite{Linde:1996gt,Mollerach:1989hu}. For a review of
curvaton scenarios see \ocite{Mazumdar:2010sa}.  For discussions of sneutrino as
curvaton see refs.  \cite{McDonald:2003xq,McDonald:2004by,Moroi:2002vx}.  For
a discussion of double sneutrino inflation see \ocite{Bi:2003ea}.}

Inflation models are typically analyzed in terms of 
the slow-roll parameters, defined as: 
\begin{equation} \label{slow}
\epsilon=\frac{m_P^2}{2}\left( \frac{V_{\chi} }{V}\right) ^{2}\,, \quad \eta
=m_P^2 \frac{V_{\chi \chi} }{V}  \,, \quad \xi ^{2} =m_P^4 \frac{V_{\chi} V_{\chi
\chi\chi} }{V^{2}}\,.  
\end{equation} 
Here $m_P\approx2.4\times10^{18}$ GeV is the reduced Planck scale, and the
subscript `$\chi$' denotes derivative with respect to the inflaton $\chi$. 

The number of e-folds during inflation is given by 
\begin{equation} \label{efold1}
N_*\equiv\ln\frac{a(t_{end})}{a(t_*)}=\int^{t_{end}}_{t_*}H\ud
t\approx\frac{1}{m_P^2}\int^{\chi_*}_{\chi_{end}}\frac{V\rm{d}\chi}{V_{\chi}}\,. 
\end{equation} 
The subscript `$_*$' implies that the values correspond to the time when the
comoving wavenumber $k_*=a_*H_*$, where $(aH)^{-1}$ is the comoving Hubble length. The
subscript `$end$' denotes the end of inflation, when
\begin{equation} \label{infend}
\epsilon_H\equiv3\frac{\dot{\chi}^2/2}{V+\dot{\chi}^2/2}=1\,.
\end{equation}

The primordial curvature perturbation $\zeta$ has a nearly scale invariant
spectrum. Assuming the only contribution to $\zeta$ is due to the fluctuations
in the inflaton field, the amplitude of the perturbation is given by 
\begin{equation} \label{perturb}
\mathcal{P}_{\zeta}\approx\frac{V}{24\pi^2 m_P^4\epsilon}\,.  
\end{equation}
The WMAP best fit value for the comoving wavenumber $k_*=0.002$ Mpc$^{-1}$ is
$\mathcal{P}_{\zeta}\approx2.4\times10^{-9}$ \cite{Komatsu:2010fb}.
The spectral index $n_s$, the tensor
to scalar ratio $r$ and the running of the spectral index
$\alpha\equiv\mathrm{d} n_s/\mathrm{d} \ln k$ are given by 
\begin{equation} \label{nsra}
n_s-1\equiv\left.\frac{\ud\ln\pz}{\ud\ln k}\right|_{k=aH} \approx - 6 \epsilon + 2 \eta \,,\quad r \approx 16 \epsilon \,,\quad
\alpha \approx  16 \epsilon \eta - 24 \epsilon^2 - 2 \xi^2\,.
\end{equation} 
Error estimates for these expressions, which are at the leading order in
slow-roll parameters, are discussed e.g. in \ocite{Lyth:1998xn}.

\subsection{Sneutrino inflation and sneutrino dominated leptogenesis} \label{sninf}

This section is a brief review of the sneutrino inflation and leptogenesis
scenario discussed in refs.  \cite{Murayama:1992ua,Ellis:2003sq}, where the
minimal supersymmetric standard model is supplemented by the superpotential
\begin{equation} \label{super} W=\frac12M_aN_aN_a+h_{a\alpha}N_aL_{\alpha}H_u\,.
\end{equation} 
Here $N_a$ ($a=1,2,3$ with the ordering $M_1< M_2< M_3$),
$L_{\alpha}$ and $H_u$ denote the superfields of the RH neutrinos, lepton
doublets and the up-type Higgs doublet, respectively. The inflaton $\chi$ is
identified with one of the three generation of sneutrinos:
$\chi\equiv\sqrt2|\widetilde{N}_i|$. In refs.
\cite{Murayama:1992ua,Ellis:2003sq} it is assumed that $i=1$, while we will be
considering the case $i\ne1$ starting from \sektion{outline}.

The inflaton mass $m_{\chi}\equiv M_i$ is fixed from $\mathcal{P}_{\zeta}$: For the potential
$V=(1/2)m_{\chi}^2\chi^2$, \eq{efold1} gives
$N_*\approx(\chi_*^2-\chi_{end}^2)/(4m_P^2)$. Numerical calculation using the
scalar field equation
\begin{equation}
\ddot{\chi}+3H\dot{\chi}+\mchi^2\chi=0\,,
\end{equation}
where $H^2=\rho/(3m_P^2)$ and $\rho=V+\dot{\chi}^2/2$ yields
$\chi_{end}\approx1.0m_P$. We define
\begin{equation} \label{def:N1}
N_+\equiv\frac{\chi_*^2}{4m_P^2}\approx N_*+\frac{\chi_{end}^2}{4m_P^2}\approx
N_*+\frac14\,.
\end{equation}
Using \eq{perturb} with $\epsilon=2m_P^2/\chi_*^2=1/(2N_+)$,
\begin{equation} \label{infmass}
m_{\chi}\approx\sqrt{6\pi^2\pz}\cdot\frac{m_P}{N_+}\,,
\end{equation}
yielding $\mchi\approx1.6\times10^{13}$ GeV for
$N_+\approx55$. From eqs. (\ref{slow}) and (\ref{def:N1}),
\begin{equation} \label{nralpha}
n_s-1\approx - \frac{8m_P^2}{\chi_*^2}\approx-\frac{2}{N_+}\,,~ 
r \approx \frac{32m_P^2}{\chi_*^2}\approx\frac{8}{N_+}\,, ~
\alpha\approx-\frac{32m_P^4}{\chi_*^4}\approx\frac{n_s-1}{N_+}\,.  
\end{equation} 

The inflaton-sneutrino decay width is given by
\begin{equation} \label{decay1}
\Gamma_{\chi}=\frac{(hh^{\dagger})_{ii}}{4\pi}\mchi\,,
\end{equation}
where $h$ is the neutrino
Yukawa couplings matrix. Defining the reheat temperature $T_r$ as the
temperature when $H=H_{reh}\equiv\gchi/2$, it is given by
\begin{equation} \label{reheat1}
T_r\approx\left(\frac{45}{\pi^2g_*}\right)^{1/4}\sqrt{\Gamma_{\chi}m_P}
\approx0.3\sqrt{\Gamma_{\chi}m_P}\,,
\end{equation}
where we have taken the relativistic degrees of freedom $g_*\approx200$.

The gravitino constraint on $T_r$ depends strongly on the gravitino mass if it
is unstable, ranging from $10^6$--$10^9$ GeV for $m_{3/2}\sim1$--10
TeV.\footnote{For the case of stable gravitinos the constraints also depend on
the next-to-lightest supersymmetric particle. For a recent analysis see
\ocite{Kawasaki:2008qe}.} Hereafter we take the gravitino constraint to be
$T_r\lesssim10^9$ GeV, which implies $(hh^{\dagger})_{ii}\lesssim10^{-11}$. As a
consequence of the suppressed Yukawa couplings, the RH neutrino superpartner of
the inflaton decouples from the see-saw mechanism.\footnote{The decoupling of a
RH neutrino can help to reconcile large mixing angles with hierarchical light
neutrino masses \cite{King:2003jb}. The patterns of lepton flavor violating
decays associated with the decoupling assumption have been discussed in ref.
\cite{Chankowski:2004jc}.} Since only the decoupled RH neutrino
contributes to the mass of the lightest left-handed neutrino, this mass is
predicted to be extremely small: 
\begin{equation} \label{mnu}
m_{\nu_1}\le\widetilde{m}_i\equiv\frac{(hh^{\dagger})_{ii}\langle
H_u^0\rangle^2}{m_{\chi}}
\approx\left(\frac{T_r}{m_{\chi}}\right)^2~2\times10^{-3}~\rm{eV}\lesssim10^{-11}~\rm{eV}\,.  
\end{equation}

The inflaton-sneutrino decays lead to a lepton asymmetry, given by \cite{Hamaguchi:2001gw}
\begin{equation} \label{asym}
Y_L\equiv\frac{n_L}{s}\approx\frac{3T_r}{4\mchi}\epsilon\lesssim1.5\times10^{-10}\frac{T_r}{10^6{\rm~GeV}}\,,
\end{equation}
where $\epsilon$ is the CP asymmetry in sneutrino decay and we have used \cite{Davidson:2002qv}
\begin{equation} \label{davidson}
\epsilon\lesssim\frac{3}{8\pi}\frac{\mchi m_{atm}}{\langle
H_u^0\rangle^2}\,.
\end{equation}
The final baryon asymmetry per entropy density due to the sphaleron processes at
equilibrium above the electroweak scale is given by $Y_B\approx Y_L/3$
\cite{Kuzmin:1985mm}. The observed baryon asymmetry of the Universe (BAU)
corresponds to $Y_B=(8.8\pm0.2)\times10^{-11}$ \cite{Komatsu:2010fb}. Thus, in
this example of non-thermal leptogenesis, the BAU can be obtained with
$T_r\gtrsim2\times10^6$ GeV. 

The only free parameter relevant to the CMB observables is $N_*$ which depends
on the reheat temperature $T_r$ logarithmically, see \subsektion{sec:zeta}.
Taking leptogenesis and gravitino constraints into account, $T_r$ is determined
to within three orders of magnitude, which corresponds to an uncertainty in
$N_*$ of just about two e-folds. As a result, the predictions for the CMB
observables are quite precise.  Using \eq{nralpha} and calculating $N_*$ as
discussed in \subsektion{sec:zeta}, we obtain $n_s=0.963(0.964)$,
$r=0.148(0.143)$ and $\alpha=-7\times10^{-4}(-6\times10^{-4})$ for
$T_r=2\times10^6(10^9)$ GeV. 

\section{Outline of the inflaton and curvaton scenario with sneutrinos} \label{outline}

In the sneutrino inflation and leptogenesis scenario summarized in
\sektion{sninf}, it was assumed that the inflaton $\chi$ has suppressed Yukawa
couplings to satisfy the gravitino constraint. We now consider an alternative
situation involving the inflaton $\chi$ and a late-decaying field $\phi$,
which are identified with different generation RH neutrino superpartners
$\widetilde{N}_i$ and $\widetilde{N}_j$, respectively. With the superpotential
given by \eq{super}, both fields $\chi$ and $\phi$ have quadratic potentials,
and it is assumed that $\mphi\ll\mchi$.\footnote{The scalar potential is
generally altered by supergravity corrections, for a review see \ocite{Lyth:1998xn}.
It is nevertheless also possible for scalar fields to have quadratic potentials
up to field values much greater than $m_P$ in supergravity, with specific non-minimal
K\"ahler potentials \cite{Linde:2005ht} or with a shift-symmetric K\"ahler potential
\cite{Kawasaki:2000yn}.} The lighter,
late-decaying $\phi$ field can eventually dominate the energy density of the
Universe so that its decay dilutes the gravitinos produced earlier and generates
the matter asymmetry \cite{Hamaguchi:2001gw}.

\begin{figure}[t!] 
\includegraphics[width=1.0\textwidth]{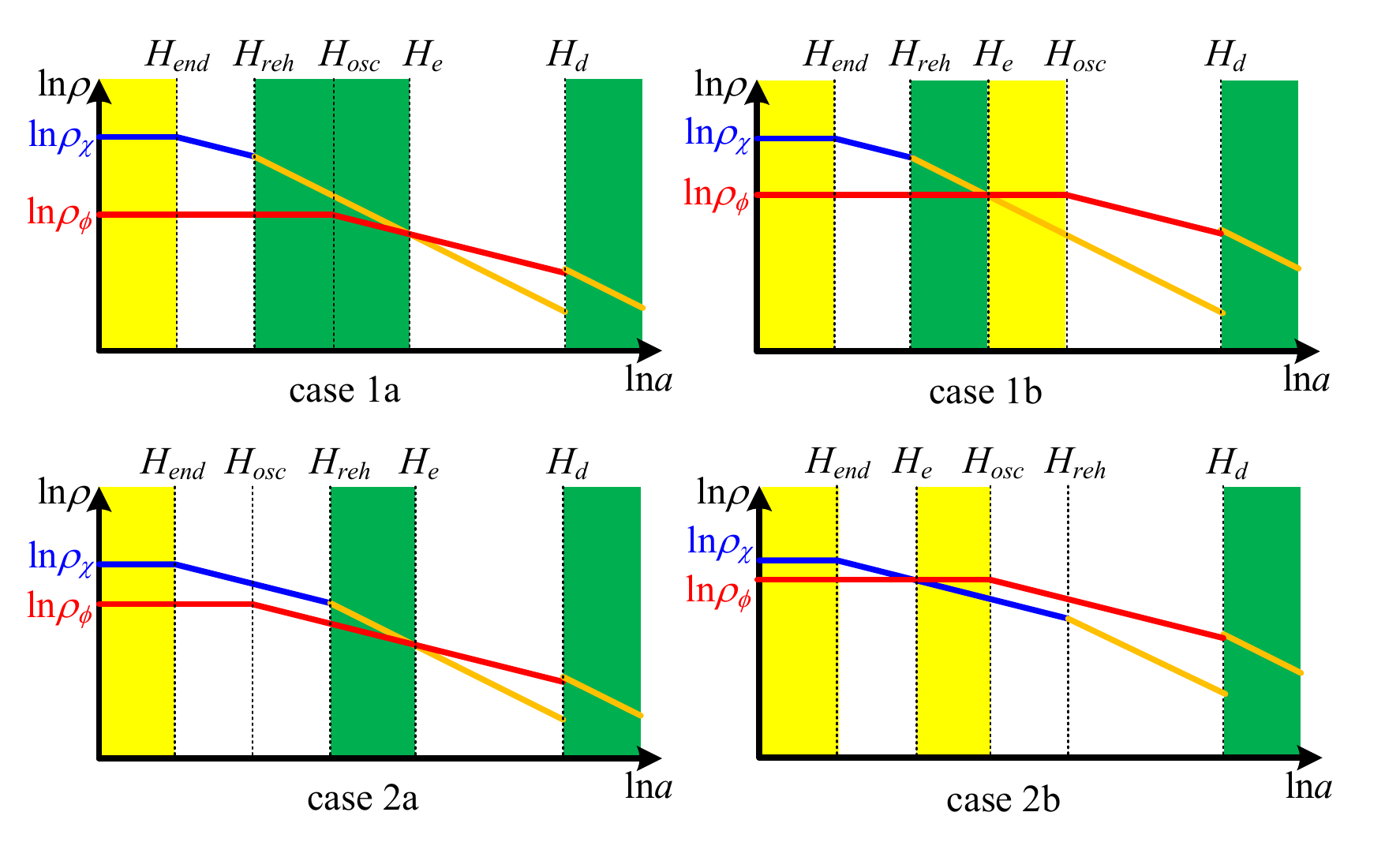}
\caption{Possible thermal histories. Inflationary epochs and radiation dominated
epochs are shaded yellow (light) and green (dark), respectively. In the unshaded regions
oscillations of either field dominate, corresponding to matter dominated
equation of state. The orange (light) segments indicate the energy density in
radiation. Case 1 and case 2 correspond to $\mphi<\gchi$ and
$\mphi>\gchi$ respectively, with a: $\phi_*\lesssim m_P$, 
b: $\phi_*\gtrsim m_P$.} \label{model} 
\end{figure}

We sketch possible thermal histories from the end of inflation until the decay
of the $\phi$ field in \fig{model}. Using \eq{infend} and $\chi_{end}\approx
m_P$, the end of inflation corresponds to
$H=H_{end}\equiv\mchi\chi_{end}/(2m_P)\approx\mchi/2$. The $\chi$ field then
starts oscillating corresponding to a matter dominated equation of state for
$\chi^2$ potential. The $\chi$ field decays when $H\sim H_{reh}\equiv\gchi/2$,
and the $\phi$ field starts oscillating when $H\sim H_{osc}\equiv\mphi/2$. We
refer to the cases $\mphi<\gchi$ and $\mphi>\gchi$ as case 1 and case 2
respectively. The energy density of the $\phi$ field $\rho_{\phi}$ equals half
the total energy density when $H\equiv H_e$ and the $\phi$ field dominates
afterwards. Finally, the $\phi$ field decays when $H\sim H_d\equiv\gphi/2$.

Different inflationary scenarios can occur depending on $\phi_*$, the
initial field value when the comoving wavenumber $k_*$ exits horizon during
inflation. As discussed in \sektion{sec:inf}, for $\phi_*\ll 0.1m_P$ the $\phi$
field is the curvaton while $\phi_*\sim0.1m_P$ corresponds to the mixed
inflaton-curvaton case, that is, both $\delta\chi$ and $\delta\phi$
significantly contribute to $\zeta$.  For $\phi_*\gtrsim m_P$ (case b), the $\phi$ field
dominates before it starts oscillating and leads to a second epoch of inflation, which can
start either after reheating (case 1b) or during $\chi$ oscillations (case 2b).
In either case, $\phi_*\sim10m_P$ again corresponds to the mixed
inflaton-curvaton case. For convenience we will refer to the late-decaying
$\phi$ field as the curvaton even though it only partially contributes to
$\zeta$ in general.
 
We can estimate $H_e$ by setting $\rho_{\phi}=\rho_{\chi}$ for case 2b, and
$\rho_{\phi}=\rho_{r}$ in the other cases, $\rho_{r}$ being the energy density
in radiation. The $\phi$ field stays almost constant at $\phi_*$ until
$H=H_{osc}$ for $\phi_*\lesssim m_P$ (case a). For $\phi_*\gtrsim m_P$ (case b)
it stays almost constant until $H=H_e$, and decreases to $\phi\approx m_P$ by
the time the second epoch of inflation ends at $H\approx H_{osc}$. Neglecting the
change in $\rho_{\phi}$ until $H_{osc}$ and $H_e$ for case a and case b
respectively, we obtain
\begin{equation}
{\rm for~case~a:~}H_e\sim\left(\frac{\phi_*}{m_P}\right)^4{\rm
min}(\mphi,\gchi)\,,\quad{\rm for~case~
b:~}H_e\sim\left(\frac{\phi_*}{m_P}\right)\mphi\,.
\end{equation}

\subsection{Constraints on the curvaton and inflaton parameters}
\label{constraints}

\paragraph{The curvaton initial value:}

A lower bound on $\phi_*$ follows from the condition that $\phi$ dominates
before it decays, so that the pre-existing gravitinos can be diluted. From
setting $H_e=H_d$ we obtain
\begin{equation} \label{phimin1}
{\rm for~case~a:~}\left(\frac{\phi_*}{m_P}\right)^4\gtrsim\frac{\Gamma_{\phi}}{{\rm min}(\mphi,\gchi)}
\,,\quad{\rm for~case~
b:~}\frac{\phi_*}{m_P}\gtrsim\frac{\Gamma_{\phi}}{\mphi}\,.
\end{equation}
However, for sufficient dilution of gravitinos $\phi$ must dominate much
earlier. Taking the gravitino constraint on the reheat temperature to be
$T_r\lesssim10^9$ GeV and recalling that the thermal abundance of gravitinos are
proportional to $T_r$, the required dilution factor $\Delta_{\rm
req}\sim T_r/(10^9~\rm{GeV})$. The dilution factor due to $\phi$ decay is
$\Delta\approx1.8 g_*^{1/4}\mphi(n_{\phi}/s)/\sqrt{\gphi m_P}$
\cite{Kolb:1990vq}.  Estimating this dilution factor for each thermal history
sketched in \fig{model}, lower bounds on $\phi_*$ corresponding to
$\Delta>\Delta_{\rm req}$ are obtained as follows:
\begin{eqnarray} \label{phimin2}
&{\rm
for~case~a:~}&\left(\frac{\phi_*}{m_P}\right)^4\gtrsim\frac{\Gamma_{\chi}}{{\rm min}(\mphi,\gchi)}
\cdot\frac{\Gamma_{\phi}}{1~\rm{GeV}}
\ge\frac{\Gamma_{\phi}}{1~\rm{GeV}}\,,\\ \nonumber
&{\rm
for~case~1b:~}&\left(\frac{m_P}{\phi_*}\right)^3\exp\left[\frac32\left(\left(\frac{\phi_*}{m_P}\right)^2-1\right)\right]\gtrsim\frac{\Gamma_{\chi}}{\mphi}\cdot\frac{\Gamma_{\phi}}{1~\rm{GeV}}\,,\\
\nonumber
&{\rm
for~case~2b:~}&\left(\frac{\phi_*}{m_P}\right)^4\exp\left[\frac98\left(\left(\frac{\phi_*}{m_P}\right)^2-1\right)\right]\gtrsim\frac{\Gamma_{\phi}}{1~\rm{GeV}}\,.
\end{eqnarray}

\paragraph{The curvaton decay width and mass:} 

Assuming the curvaton dominates before it decays, the reheat temperature after
the decay is obtained as in \eq{reheat1}:
$T_d\approx0.3\sqrt{\Gamma_{\phi}m_P}$. The lepton asymmetry created by the
decays of $\phi$ is given by \eq{asym} with $T_r$ and $\mchi$ replaced by $T_d$
and $\mphi$. The constraints from sneutrino dominated leptogenesis and gravitino
over-production ($2\times10^6$ GeV $\lesssim T_d\lesssim10^9$ GeV) correspond to
$10^{-5}$ GeV $\lesssim \Gamma_{\phi}\lesssim10$ GeV.

Throughout the paper we assume that the curvaton mass
$\mphi\ll\mchi$. As for a lower bound, generating the BAU via sneutrino
dominated leptogenesis requires $m_{\phi}\gtrsim T_d\gtrsim2\times10^6$ GeV.
To avoid thermalization of the condensate,
$\widetilde{m}_j\lesssim2\times10^{-3}$ eV is required
\cite{McDonald:2003xq,McDonald:2004by}. Since
\begin{equation} \label{mnu2}
m_{\nu_1}\le\widetilde{m}_j\equiv\frac{(hh^{\dagger})_{jj}\langle
H_u^0\rangle^2}{m_{\phi}}
\approx\left(\frac{T_d}{m_{\phi}}\right)^2~2\times10^{-3}~\rm{eV}\,,  
\end{equation}
we see that this condition is also satisfied for $m_{\phi}\gtrsim T_d$. Note
that part of the asymmetry can be washed out if $m_{\phi}\sim T_d$, but for
$\widetilde{m}_j\lesssim2\times10^{-3}$ eV the washout is weak and enough
asymmetry can remain \cite{Buchmuller:2004nz}. 

Another constraint on $\Gamma_{\phi}$ and $m_{\phi}$ comes from the asymmetry
created by a variation of the Affleck-Dine (AD) mechanism \cite{Affleck:1985fy},
discussed in \ocite{Allahverdi:2004ix}. The asymmetry is produced due to the
rotation of the condensate as in the AD mechanism, but the asymmetry that
survives the decay of sneutrinos depends on supersymmetry breaking as in soft
leptogenesis \cite{Grossman:2003jv}. Ref. \cite{McDonald:2004by} estimates
the asymmetry resulting from the soft supersymmetry breaking $B$-term for the sneutrino
and concludes that $m_{\phi}\gtrsim10^8$ GeV and $m_{\nu_1}\lesssim10^{-8}$ eV is
required for sneutrino dominated leptogenesis to work. On the other hand, it is
shown in \ocite{Allahverdi:2004ix} that the asymmetry from the $B$-term is negligible by
itself, and including the more important thermal effects results in an asymmetry that for
$\mphi\gtrsim T_d$ is given by 
\begin{equation} \label{bau1}
\frac{n_L}{s}\sim10^{-8}\left(\frac{\Gamma_{\phi}}{10^{-5}~\rm{GeV}}\right)\left(\frac{1~\rm{TeV}}{B}\right)\left(\frac{T_d}{m_{\phi}}
\right)\left|\exp\left(-\frac{\mphi}{2T_d}\right)-\frac{0.09T_d^2}{\mphi^2}\right|\,. 
\end{equation} 
It is thus possible for the BAU to be generated by this mechanism with
$\mphi\gtrsim T_d\gtrsim10^{5.5}$ GeV. However, baryon isocurvature
perturbations arise if the $\phi$ field starts oscillating before it dominates
(case a) \cite{Hamaguchi:2003dc,McDonald:2004by}. For these isocurvature
perturbations to remain below the observational bounds, either the asymmetry
given by \eq{bau1} should be subdominant (which requires $\mphi\gtrsim4T_d$) or
the contribution from the $\phi$ field to the curvature perturbation should be
subdominant. As discussed in \sektion{sec:inf}, keeping this contribution below
10\% requires $\phi_*\gtrsim0.3m_P$.

\paragraph{The inflaton decay width and mass:}

The inflaton decay width $\gchi$ is given by \eq{decay1}. Using \eq{mnu}, it can
be expressed as follows:
\begin{equation}
\gchi=\frac{\widetilde{m}_i\mchi^2}{4\pi\langle H_u^0\rangle^2}\sim
\left(\frac{\widetilde{m}_i}{0.05~\rm{eV}}\right)
\left(\frac{\mchi}{10^{13}~\rm{GeV}}\right)^2 10^{10}~\rm{GeV}\,.
\end{equation}
If the contribution of $\delta\phi$ to
$\zeta$ is negligible, $m_{\chi}\approx1.6\times10^{13}$ GeV as mentioned in
\sektion{sninf}. As will be explained in \sektion{sec:zeta}, $m_{\chi}$ 
depends on $\phi_*$ in the mixed inflaton-curvaton case. Applying the WMAP and
gravitino constraints on $\phi_*$ and using \eq{infmass2}, $\mchi$ is found to
vary in the range 0.9--2.2$\times10^{13}$ GeV, see \fig{myn}.

\section{Inflationary predictions} \label{sec:inf}

\subsection{Calculating the power spectrum and the number of e-folds} \label{sec:zeta}

Using the $\delta N$ formalism
\cite{Starobinsky:1986fxa,Sasaki:1995aw,Ichikawa:2008iq}, the primordial
curvature perturbation $\zeta$ can be written as
\begin{equation}
\zeta=\delta N_{tot}\approx\frac{\partial N_{tot}}{\partial\chi}\delta\chi_*
+ \frac{\partial N_{tot}}{\partial\phi}\delta\phi_*\,,
\end{equation}
where $N_{tot}$ is the number of e-folds from horizon exit (of the scale
corresponding to the comoving wavenumber $k_*$) to some final time $t_f$ well
after the curvaton has decayed, when $H=H_f\ll\gphi$. We can separate $N_{tot}$
into two parts $N_{tot}=N_*+N$, where $N_*$ is the number of e-folds from horizon exit to
the end of inflation and $N$ is the number of e-folds from the end of inflation to $t_f$.
Since $N$ does not depend on $\chi$,
\begin{equation}
\frac{\partial N_{tot}}{\partial\chi}=\frac{\partial N_*}{\partial\chi}
\approx\frac{V}{m_P^2V_{\chi}}\,,
\end{equation}
where in the last step we have used \eq{efold1}. Similarly, since $N_*$ does not
depend on $\phi$,
$\partial N_{tot}/\partial\phi=\partial N/\partial\phi\equiv\nphi$. Thus,
\begin{equation}
\zeta\approx\frac{V}{m_P^2V_{\chi}}\delta\chi_*+\nphi\delta\phi_*\,.
\end{equation}
Assuming $\delta\chi_*$ and $\delta\phi_*$ to be uncorrelated, the power spectrum of the
perturbation is then
\begin{equation}
\mathcal{P}_{\zeta}\approx\left(\frac{V^2}{m_P^4V_{\chi}^2}+\nphi^2\right)\left(\frac{H}{2\pi}\right)^2\,.
\end{equation}
Defining $y\equiv2m_P^2\nphi^2\epsilon$, this equation can be written as
\begin{equation} \label{perturb2}
\mathcal{P}_{\zeta}\approx\frac{(1+y)V}{24\pi^2 m_P^4\epsilon}\,.
\end{equation}
The contribution of the $\phi$ field to the curvature perturbation is negligible
for $y\ll1$ whereas $y\gg1$ corresponds to the curvaton limit. For $y\sim1$ the
mixed inflaton-curvaton case is realized. From eqs. (\ref{perturb2}) and
(\ref{def:N1}) the inflaton mass is given as
\begin{equation} \label{infmass2}
m_{\chi}\approx\sqrt{\frac{6\pi^2\pz}{1+y}}\cdot\frac{m_P}{N_+}\,.
\end{equation}

Since $y$ depends on $N_*$ as well as $\nphi$, we now discuss how to calculate the
e-fold numbers $N_*$ and $N$.  Using the definition of
$N_*\equiv\ln(a_{end}/a_*)$ where $k_*=a_*H_*$ at horizon exit, we can relate
$N_*$ to the current scale factor $a_0$ and Hubble parameter $H_0$ as follows:
\begin{equation}
N_*=-\ln\frac{k_*}{a_0H_0}+\ln\frac{a_{end}}{a_f}+\ln\frac{a_f}{a_0}+\ln\frac{H_*}{H_0}\,.
\end{equation}
In this expression the first term at the right hand side is fixed from
$k_*=0.002$ Mpc$^{-1}$, the second term is $-N$ and the last term can be
expressed in terms of $\mchi$ and $N_*$. For the third term note that any
significant entropy production after $t_f$ would dilute the $B-L$ asymmetry
created by the decays of $\phi$. Therefore assuming no significant entropy
production,
\begin{equation}
\ln\frac{a_f}{a_0}=\frac13\ln\frac{s_0}{s_f}=\frac13\ln\frac{g_{*s0}T_0^3}{g_{*s}T_f^3}\,.
\end{equation}
Using $\rho_f=3H_f^2m_P^2=(\pi^2/30)g_*T_f^4$ and taking the relativistic degrees
of freedom $g_*=g_{*s}=200$ we obtain
\begin{equation} \label{nstar}
N_*-\frac12\ln N_*\approx64.3+\ln\frac{\mchi}{m_P}+\frac12\ln\frac{m_P}{H_f}-N\,.
\end{equation}

To estimate $N$, we can add the number of e-folds in each matter dominated,
radiation dominated or inflationary epoch between the end of inflation $t_{end}$
corresponding to $H=H_{end}$ and the final time $t_f$ corresponding to $H=H_f$
assuming the transitions between the epochs to be sudden. For instance, in the
case of sneutrino inflation with no late-decaying $\phi$ field, the universe has
a matter dominated equation of state (for $\chi^2$ potential) between $H_{end}$
and $H_{reh}$, and radiation dominates after $H=H_{reh}$. Therefore
$N\approx(2/3)\ln(H_{end}/H_{reh})+(1/2)\ln(H_{reh}/H_f)$. Using this with eqs.
(\ref{nstar}) and (\ref{nralpha}),   we obtain the values of $n_s$, $r$ and
$\alpha$ given in \subsektion{sninf}.

In the presence of the late-decaying $\phi$ field, four possible thermal
histories were discussed in \sektion{outline}. For case 1a, there are
alternating matter-radiation-matter-radiation dominated epochs between $H_{end}$
and $H_f$ (see \fig{model}) so that
\begin{equation} 
N\approx\frac23\ln\frac{H_{end}}{H_{reh}}+\frac12\ln\frac{H_{reh}}{H_{e}}
+\frac23\ln\frac{H_{e}}{H_{d}}+\frac12\ln\frac{H_{d}}{H_{f}}\,.
\end{equation}
Using $H_{end}\approx\mchi/2$, $H_{reh}\equiv\gchi/2$,
$H_e\sim(\phi_*/m_P)^4\mphi$ and $H_d\equiv\gphi/2$, the
result shown in table \ref{tab:N} is obtained. For case 1b, the $\phi$ field
dominates before it starts oscillating, leading to a second inflationary epoch
between $H_e$ and $H_{osc}$ lasting $\approx\phi_*^2/(4m_P^2)$ e-folds. The
results are similar for case 2. Note that both for case 1 and case 2,
$\nphi\approx2/(3\phi_*)$ in the limit $\phi_*\ll m_P$, and
$\nphi\approx\phi_*/(2m_P^2)$ in the limit $\phi_*\gg m_P$. These
results were also obtained in refs.
\cite{Langlois:2004nn,Moroi:2005kz,Ichikawa:2008iq},
where only case 1 was considered.

\begin{table}[t!]
\begin{tabular}{lll}
\toprule
 & Case a ($\phi_*\lesssim m_P$) & Case b ($\phi_*\gtrsim m_P$) \\
\cmidrule(l){2-3}
Case 1 ($\mphi<\gchi$) &
$N\approx\frac23\ln\frac{\phi_*}{m_P}+\frac16\ln\frac{\mchi^4\mphi}{\gchi\gphi
H_f^3}$ & $N\approx\frac{\phi_*^2}{4m_P^2}-\frac12\ln\frac{\phi_*}{m_P}+\frac16\ln\frac{\mchi^4\mphi}{\gchi\gphi
H_f^3}$ 
 \\
\cmidrule(l){2-3}
Case 2 ($\mphi>\gchi$) & $N\approx\frac23\ln\frac{\phi_*}{m_P}+\frac16\ln\frac{\mchi^4}{\gphi
H_f^3}$ & $N\approx\frac{\phi_*^2}{4m_P^2}-\frac23\ln\frac{\phi_*}{m_P}+\frac16\ln\frac{\mchi^4}{\gphi
H_f^3}$
 \\
\bottomrule
\end{tabular}
\caption{The approximate number of e-folds $N$ from the end of inflation
$t_{end}$ to a final time $t_f$, for the four
thermal histories discussed in \sektion{outline}.}
\label{tab:N}
\end{table}

For a more accurate calculation, we numerically solve the following background
equations, from $t_{end}$ to $t_f\gg\gphi^{-1}$: 
\begin{equation} \label{back}
\dot{\rho_{\chi}}+3H\rho_{\chi}  =  - \gchi\rho_{\chi}\,,\;\;
\ddot{\phi}+(3H+\gphi)\dot{\phi}+\mphi^2\phi  =  0 \,,\;\; 
\dot{\rho_{r}}+4H\rho_r  =  \gchi\rho_{\chi}+\gphi\rho_{\phi}\,,
\end{equation}
where $\rho_{\phi}=\dot{\phi}^2/2+(1/2)\mphi^2\phi^2$ and
$H^2=(\rho_{\chi}+\rho_{\phi}+\rho_{r})/(3m_P^2)$. The initial conditions are
taken as follows:
\begin{equation} \label{ini}
\rho_{\chi}(t_{end})=\frac34\mchi^2m_P^2\,,\quad
\rho_r(t_{end})=0\,,\quad \phi(t_{end})=\phi_*\,,\quad
\dot{\phi}(t_{end})=-\frac{\mphi^2\phi_*}{3H_{end}}\,.
\end{equation}
To calculate $N$ and $\nphi$, we first estimate $\mchi$ using eqs.
(\ref{infmass2}), (\ref{nstar}) and the approximate expressions for $N$ given in
table \ref{tab:N}.  Using eqs. (\ref{back}) and (\ref{ini}) we evaluate $N$ numerically
for different values of $\phi_*$ and interpolate to obtain $N$ as a function of $\phi_*$.
We then recalculate $\mchi$ and iterate this procedure.

\subsection{Observational quantities} \label{sec:ns}

In the sneutrino inflaton-curvaton scenario we have outlined, we have assumed
that the late-decaying curvaton dominates the Universe before it decays.
Non-Gaussianities are then not expected to be large \cite{Lyth:2002my,Ichikawa:2008iq}. As
discussed in \sektion{constraints}, there could be isocurvature perturbations at
an observable level if both $\phi_*\lesssim0.3m_P$ and the matter asymmetry
mostly originates from a variation of the AD mechanism. Since this is not a
general prediction, our discussion of observational quantities will focus on the
usual parameters $n_s$ and $r$. We will also briefly comment on $\alpha$.

In the presence of the late-decaying $\phi$ field, 
the following expressions are obtained using \eq{perturb2}
\cite{Langlois:2004nn,Ichikawa:2008iq}:
\begin{equation} 
n_s - 1 \approx - 2 \epsilon +\frac{ 2 \eta-4\epsilon}{1+y} \,,\quad 
r \approx \frac{16 \epsilon}{1+y} \,,\quad
\alpha \approx 4\epsilon(\eta-2\epsilon)+\frac{12 \epsilon \eta - 16 \epsilon^2 - 2
\xi^2}{1+y}\,.
\end{equation}

For inflation with $\chi^2$ potential, $\xi^2=0$ and from \eq{def:N1},
$\epsilon=\eta=1/(2N_+)$. In terms of $N_+\approx N_*+1/4$ and $y$ we have
\begin{equation} \label{nsra2}
n_s - 1 \approx - \frac{1}{N_+}\left(\frac{2+y}{1+y}\right)  \,,\quad 
r \approx \frac{8}{N_+(1+y)} \,,\quad
\alpha \approx \frac{n_s-1}{N_+}\,.
\end{equation}

As discussed in \subsektion{sec:zeta}, the values of the decay widths $\gchi$,
$\gphi$ and the curvaton mass $\mphi$ only have a small (logarithmic) effect on
the number of e-folds. Thus, values of $n_s$, $r$ and $\alpha$ depend
mostly on the initial value $\phi_*$. For a qualitative discussion of the
results it is convenient to define another parameter $N_0\equiv
N_++\phi_*^2/(4m_P^2)$. Since the $\phi$ field leads to a second inflationary
epoch lasting $\approx\phi_*^2/(4m_P^2)$ e-folds for $\phi_*\gtrsim m_P$, $N_0$
is approximately the total number of e-folds during the two inflationary epochs.

For $\phi_*\lesssim m_P$ (case a), $N_0\approx N_+\approx55$.
Using $\nphi\approx2/(3\phi_*)$ and $y=m_P^2\nphi^2/N_+$, we see that the curvaton
limit $y\gg1$ applies if $\phi_*/m_P\ll2/(3\sqrt{N_+})\approx0.09$. In this limit
\begin{equation} \label{nsra3}
n_s-1\approx-\frac{1}{N_+}\,,\quad
r\approx\frac{8}{yN_+}\approx\frac{18\phi_*^2}{m_P^2}\,,\quad
\alpha\approx-\frac{1}{N_+^2}\,.
\end{equation}
For $\phi_*\sim m_P$, the standard predictions of inflation with $\chi^2$
potential -- given by \eq{nralpha} -- are recovered since $y\ll1$.  Finally, for
$\phi_*\gtrsim m_P$ (case b), $N_+\approx N_0-\phi_*^2/(4m_P^2)$ while $N_0$
remains approximately constant. Using $\nphi\approx\phi_*/(2m_P^2)$ we obtain
$y\approx\phi_*^2/(4m_P^2N_+)$ and $yN_+\approx N_0-N_+$. Expressing \eq{nsra2}
in terms of $N_0$ and $N_+$ yields
\begin{equation} \label{nsra4}
n_s - 1\approx - \frac{1}{N_+}-\frac{1}{N_0}  \,,\quad 
r \approx\frac{8}{N_0} \,,\quad
\alpha \approx\frac{n_s-1}{N_+}\,.
\end{equation}
It follows from these expressions that the spectrum is more red-tilted for
larger values of $\phi_*$, whereas the tensor to scalar ratio $r$ remains
essentially constant. Note that while $\delta\phi$ can partially contribute to
$\zeta$, the curvaton limit $y\gg1$ is not possible for case b, since
cosmological scales do not exit the horizon during the initial epoch of
inflation driven by $\chi$ if $\phi_*\gtrsim14m_P$.\footnote{The curvaton
potential must be different from quadratic for the inflating curvaton scenario
to be realized \cite{Dimopoulos:2011gb}.}

We now consider how the results depend on the decay widths $\gchi$, $\gphi$ and
the curvaton mass $\mphi$. $N_0$ increases with $\gphi$ and also if
$\gchi>\mphi$, since the radiation dominated epochs last longer. Following the
discussion of the gravitino constraint in \sektion{outline}, for the high $N_0$
case we take $\gphi=10$ GeV corresponding to a reheat temperature $T_d\sim10^9$
GeV.  We also take $\gchi=10^{12}$ GeV and $\mphi=10^{11}$ GeV corresponding to
case 1.\footnote{Since $\mphi\sim100T_d$, the asymmetry given by \eq{bau1} is
suppressed to an order of magnitude compatible with the observed BAU.} For the
low $N_0$ case we take $\gphi=10^{-5}$ GeV, since it is difficult to obtain
sufficient matter asymmetry for lower values. Minimizing $N_0$ requires
$\gchi<\mphi$, to be specific we take $\gchi=10^9$ GeV and $\mphi=10^{11}$ GeV,
corresponding to case 2. From table \ref{tab:N} we see that there is about three
e-folds difference between the two cases. 

\begin{figure}[t!]
\begin{center}
\includegraphics[width=14cm]{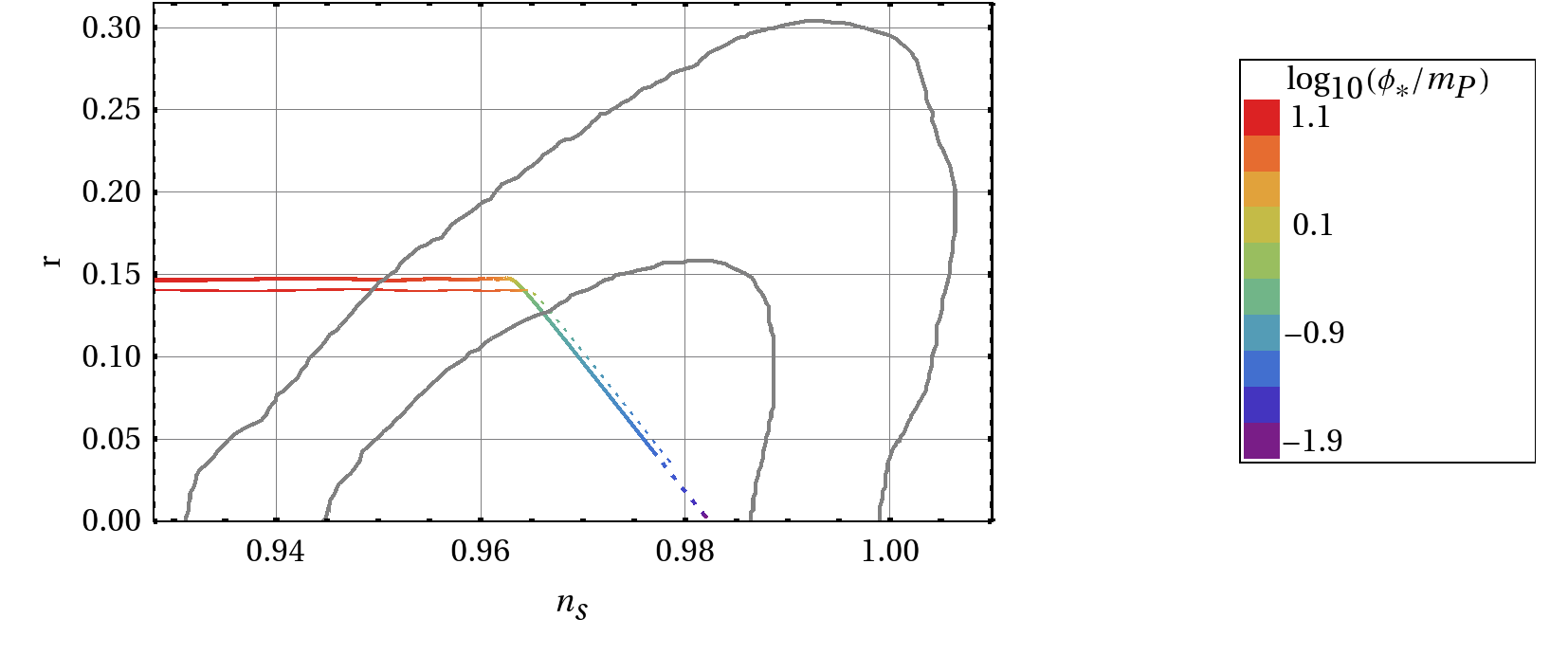} \\
\includegraphics[width=7cm]{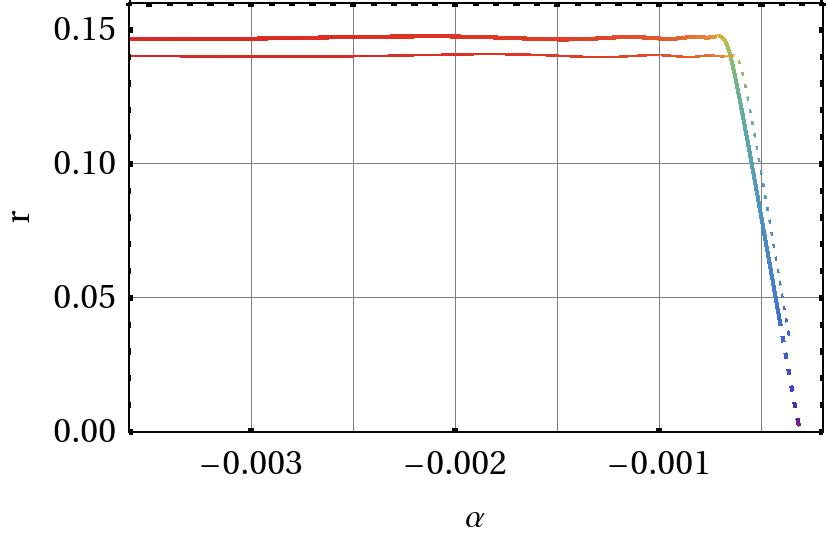}
\includegraphics[width=7cm]{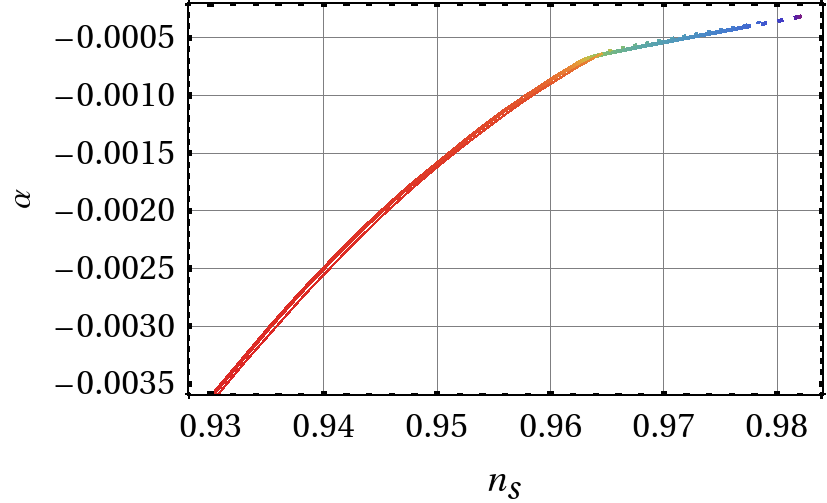}
\end{center}
\caption{$r$ vs $n_s$, $r$ vs $\alpha$ and $\alpha$ vs $n_s$ for the low
$N_0$ case (thick curves) and the high $N_0$ case (thin curves). The gravitino
constraint \eq{phimin2} is not satisfied in the dashed segments. The $r$ vs
$n_s$ plot also displays the WMAP contours as obtained with the WMAP
Cosmological Parameter Plotter (see ref. \cite{wmapweb}, model:
lcdm+sz+lens+tens, data: wmap7+bao+h0).}
\label{fig:nra}
\end{figure}

\begin{figure}[t!]
\begin{center}
\includegraphics{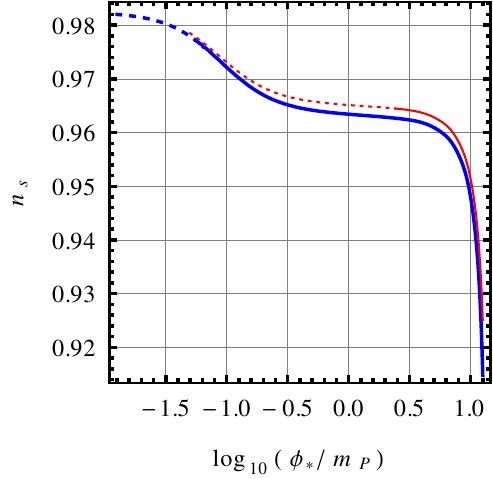}
\includegraphics{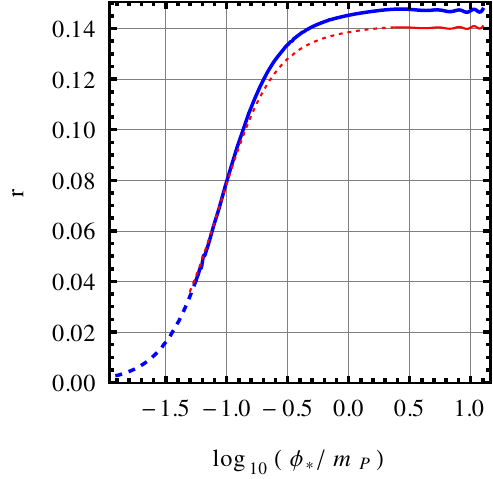}
\includegraphics{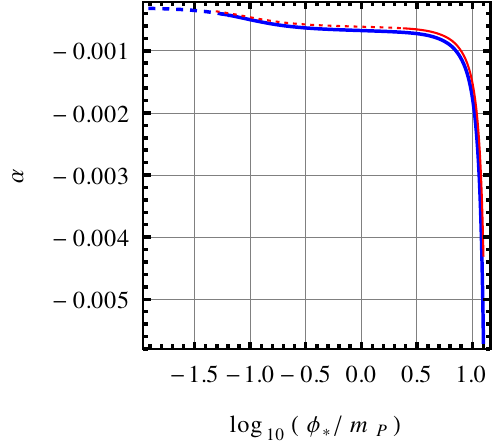}
 \\
\includegraphics{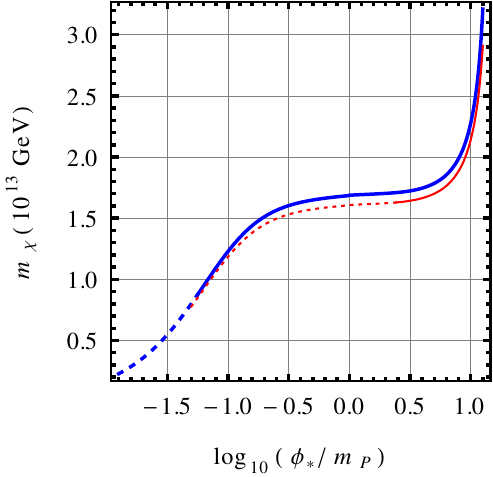}
\includegraphics{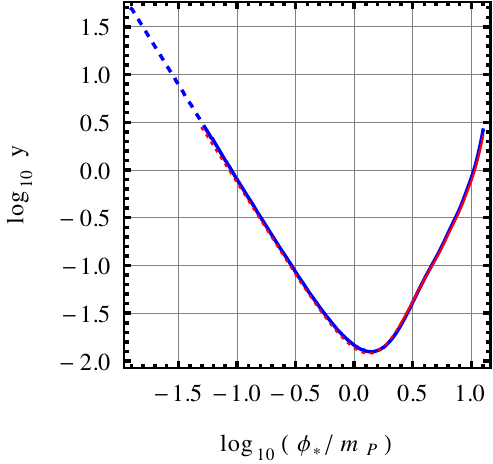}
\includegraphics{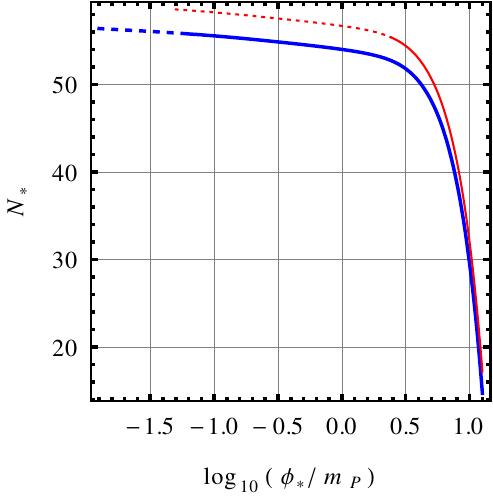}
\end{center}
\caption{$n_s$, $r$, $\alpha$, $\mchi$, $y$ and $N_*$ vs $\phi_*$ for the low
$N_0$ case (thick blue curves) and the high $N_0$ case (thin red curves). The gravitino
constraint \eq{phimin2} is not satisfied in the dashed segments.}
\label{myn}
\end{figure}

Numerical results for the two cases are displayed in \fig{fig:nra} and
\fig{myn}.  Note that although the tensor to scalar ratio $r$ becomes negligible
in the small $\phi_*$ limit, the leptogenesis constraint $\gphi\gtrsim10^{-5}$
GeV together with the gravitino constraint \eq{phimin2} requires
$\phi_*\gtrsim10^{-5/4}m_P$.  This implies $r\gtrsim0.04$, which can be observed
by the Planck satellite \cite{planck,Planck:2006aa}. For the high $N_0$ case the
gravitino constraint corresponds to $\phi_*\gtrsim2.4m_P$.

To summarize, in the presence of the late-decaying $\phi$ field, the predictions
for $n_s$ and $r$ can be distinguished from the predictions of inflation with
$\chi^2$ potential for values of $\phi_*$ which are either $\sim0.1m_P$ or
$\sim10m_P$, corresponding to the mixed inflaton-curvaton case. (The curvaton
limit $\phi_*\ll0.1m_P$ is disfavored by the gravitino constraint.) For
$\phi_*\sim0.1m_P$, $n_s\approx0.97$ and $r<0.1$ yet large enough to be
observable. For $\phi_*\sim10m_P$, $n_s\lesssim0.96$ and $r=0.14$--$0.15$.

In the mixed inflaton-curvaton case the single field consistency
relation $n_T=-r/8$ gets modified to \cite{Langlois:2004nn}
\begin{equation}
n_T=-\frac{(1+y)r}{8}\,,
\end{equation}
implying a more red-tilted tensor spectrum. Although testing the single-field consistency
relation and observing a deviation from it is beyond the forecasted accuracy of the
Planck mission, it might be possible with future CMB
observations provided $r\gtrsim0.1$ \cite{Song:2003ca}.

The running of the spectral index $\alpha$ remains small in general, although it
is enhanced for $\phi_*\gg m_P$. The WMAP constraints on $n_s$ and $r$ are
satisfied for $\phi_*\lesssim9.6m_P$, which implies
$\alpha\gtrsim-1.6\times10^{-3}$.  This running could perhaps be observed by
future galaxy surveys and 21 cm experiments \cite{Adshead:2010mc}. 

\section{Conclusion and discussion}

In the simple yet successful scenario of sneutrino inflation based on $\chi^2$
potential, the sneutrino driving inflation can also generate the matter
asymmetry via non-thermal leptogenesis \cite{Murayama:1992ua,Ellis:2003sq}. We
have considered a variation of this scenario, by assuming that the sneutrino
driving inflation is distinct from the late-decaying sneutrino. The lighter,
late-decaying sneutrino can partially contribute to the curvature perturbation
and alter the predictions for the CMB observables. The late decay of this
sneutrino also dilutes the gravitinos produced earlier and generates the
matter asymmetry.

In this mixed inflaton-curvaton scenario, since the late-decaying sneutrino is
lighter its Yukawa couplings and therefore the lightest neutrino mass
$m_{\nu_1}$ do not have to be as small as in the original version of sneutrino
inflation, which requires $m_{\nu_1}\lesssim10^{-11}$ eV  to satisfy the
gravitino constraint. However $m_{\nu_1}$ is still constrained to be
$\lesssim2\times10^{-3}$ eV (see \sektion{constraints}). 

Considering the CMB observables the mixed inflaton-curvaton scenario is less
predictive, since $n_s$ and $r$ depend on $\phi_*$, the initial value of the
late-decaying sneutrino. For $\phi_*\sim m_P$ the predictions are the same as
inflation with $\chi^2$ potential.  However, the predictions change
significantly for values of $\phi_*$ which are either $\sim0.1m_P$ or
$\sim10m_P$. In the first case $n_s\approx0.97$ and $r<0.1$ yet large enough to
be observable, whereas in the latter case $n_s\lesssim0.96$ and
$r=0.14$--$0.15$.

The original version of sneutrino inflation fixes the mass of one RH neutrino
but provides no constraints on the other two, except that they should not
decouple from the see-saw mechanism. Whereas in the mixed inflaton-curvaton
scenario the inflaton sneutrino is distinct from the late-decaying sneutrino, so
it is possible to determine the mass of the former using the CMB observables and
put a lower limit on the mass of the latter from the observed matter asymmetry.
If future CMB data remain consistent with this scenario, it would be of interest
to embed it in a more predictive model connecting CMB observables to low-energy
leptonic observables.

\bibliography{sncurv3}

\end{document}